\documentclass[letterpaper,10pt]{article}
\PassOptionsToPackage{hyphens}{url}
\usepackage[utf8]{inputenc}
\usepackage[T1]{fontenc}
\usepackage{graphicx}
\usepackage{longtable}
\usepackage{wrapfig}
\usepackage{rotating}
\usepackage[normalem]{ulem}
\usepackage{amsmath}
\usepackage{amssymb}
\usepackage{capt-of}
\usepackage{hyperref}
\usepackage{aaai22}
\usepackage{times}  
\usepackage{helvet}  
\usepackage{courier}  
\usepackage[hyphens]{url}  
\usepackage{graphicx} 
\usepackage{natbib}  
\usepackage{caption} 
\DeclareCaptionStyle{ruled}{labelfont=normalfont,labelsep=colon,strut=off} 
\usepackage{algorithm}
\usepackage{algorithmic}

\usepackage{newfloat}
\usepackage{listings}
\floatstyle{ruled}
\newfloat{listing}{tb}{lst}{}
\floatname{listing}{Listing}

\usepackage{color}
    \definecolor{MyBrown}{rgb}{0.3,0,0}
    \definecolor{MyBlue}{rgb}{.2,.2,.8}  
    \definecolor{MyRed}{rgb}{0.4,0,0.1}
    \definecolor{MyGreen}{rgb}{0,0.4,0}
\hypersetup{breaklinks=true}
    \hypersetup{
                bookmarksnumbered = true,
                bookmarksopen = true, 
                bookmarksopenlevel = 2, 
                pdfpagemode = UseOutlines,
                colorlinks = true,
                linkcolor = MyRed,
                citecolor = MyBlue,
                filecolor = MyBrown,
                urlcolor = MyGreen}

\author{Keng-Chi Chang}
\affiliations{Department of Political Science\\ University of California, San Diego\\ kechang@ucsd.edu, kengchichang.com}
\date{\today}
\title{Mapping Visual Themes among Authentic and Coordinated Memes}
\begin{document}

\maketitle
\begin{abstract}
What distinguishes authentic memes from those created by state actors?
I utilize a self-supervised vision model, DeepCluster, to learn low-dimensional visual embeddings of memes and apply K-means to jointly cluster authentic and coordinated memes without additional inputs.
I find that authentic and coordinated memes share a large fraction of visual themes but with varying degrees.
Coordinated memes from Russian IRA accounts promote more themes around celebrities, quotes, screenshots, military, and gender.
Authentic Reddit memes include more themes with comics and movie characters.
A simple logistic regression on the low-dimensional embeddings can discern IRA memes from Reddit memes with an out-sample testing accuracy of 0.84.
\end{abstract}

\section*{Introduction}
\label{sec:orge5dbb5b}
\noindent Visual memes (broadly defined as images-with-text) are everywhere on social media; a large fraction is political.
According to a panel of 490K Twitter users with voter registration, 19\% of their tweets are classified as memes, and 30\% of the memes are politically relevant \citep{du_2020_UnderstandingVisualMemes}.
Another study on political misinformation in Indian WhatsApp groups finds that 30\% of the visual misinformation are memes \citep{garimella_2020_ImagesMisinformationPolitical}.
There are legitimate concerns around state-linked online information operations affecting political behavior, but most studies to date do not leverage the wealth of data in images.

This project aims to document what kinds of visual frames are commonly promoted by state actors compared to generic, authentic memes promoted by regular non-state users.
I use a large sample of data from Russian IRA accounts released by Twitter and collect a large sample of authentic memes from \texttt{r/memes} on Reddit.
I feed a balanced sample of both coordinated state-linked memes and authentic memes (memes promoted by regular users) into a self-supervised vision model \citep{caron_2019_DeepClusteringUnsupervised} to learn the lower-dimensional representations for each meme.
I then apply the standard K-means clustering algorithm to the representations to find the clusters of memes.
I find that coordinated and authentic memes differ in the visual themes and that a simple logistic regression on the lower-dimensional representations can achieve reasonable accuracy in predicting coordinated vs. authentic memes (on the test set, AUC 0.91, accuracy 0.84, F\textsubscript{1}-score 0.84).

Compared with similar methods relying on multimodal neural networks \citep{beskow_2020_EvolutionPoliticalMemes,du_2020_UnderstandingVisualMemes}, Bag-of-Visual-Words (BOVW), or Perceptual hashing (pHash) \citep{zannettou_2018_OriginsMemesMeans,zannettou_2019_CharacterizingUseImages}, this transfer learning framework does not rely on extensive tagging (cf. multimodal models), does not only learn on local visual features (cf. BOVW), and does not require memes to be nearly identical (cf. pHash).

\section*{Prior Works and Limitations}
\label{sec:orgb7713ee}

Twitter released a ground truth dataset of state-linked operations, including the 1.8M images posted by the accounts controlled by the Russian Internet Research Agency (IRA) during the 2016 US Presidential election.
Qualitative studies pointed out that IRA employees are assessed for meme-making capabilities \citep{diresta_2021_InHouseVsOutsourced}.
Other studies used \emph{textual data} from IRA found asymmetric flooding of \emph{entertainment}, not necesarrily \emph{politics}, as a strategy \citep{cirone_2022_AsymmetricFloodingTool}, and that textual content is a reasonable predictor for state-linked campaigns \citep{alizadeh_2020_ContentbasedFeaturesPredict}.
Previous work, in a similar effort, also documented the spread of IRA memes online using Perceptual hashing (pHash) of images \citep{zannettou_2019_CharacterizingUseImages}.
However, there is a lack of systematic understanding of the amplification of visual themes by state-linked actors compared to organic ones.
\section*{Methodology}
\label{sec:org084f2ea}

This project has the following steps, which will be explained in subsections.
\begin{enumerate}
\item Collect state-linked images, organic memes, and non-meme image-with-text data (as negative samples).
\item Classify state-linked images into memes vs. non-memes.
\item Extracting embeddings of visual feature \emph{jointly} for both coordinated and authentic memes.
\item Cluster memes based on the vectors of embeddings.
\item Label the clusters and compare the difference in proportions between coordinated and authentic memes.
\item Train a simple baseline model based on visual embeddings to distinguish coordinated and authentic memes.
\end{enumerate}

\subsection*{Data collection}
\label{sec:orge3c83a2}

The primary coordinated data are the images shared by IRA on Twitter.
Other than the dataset from Twitter,
I collected 26K generic memes collected from the \texttt{r/meme} subreddit and 15K non-meme image-with-text data (COCO-text, \cite{veit_2016_COCOTextDatasetBenchmark}) as negative samples for training meme vs. non-meme classifier.
The reason for choosing non-meme image-with-text data as negative samples for training is that we would not want models simply picking up textual features in images and classifying images into with vs. without text.
We contend there there can still be coordinated memes in the \texttt{r/meme} subreddit, but the percentage should be low and is ouside of the scope of this project.

\subsection*{Classify images into memes}
\label{sec:org77d9ad5}

\begin{figure}[!t]
\centering
\includegraphics[width=.35\textwidth]{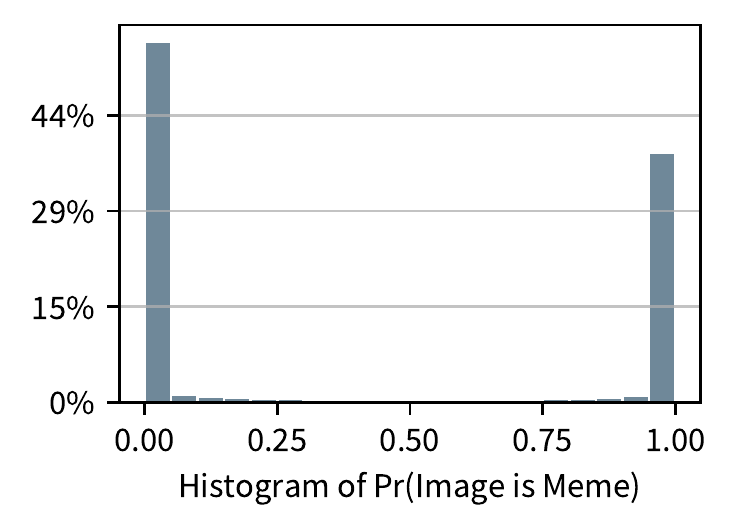}
\caption{\label{fig:classifier-percentage}Distribution of predicted probability of memes for IRA images}
\end{figure}

I trained a state-of-the-art deep learning classifier based on ResNet-50 to classify the images shared by IRA into memes vs. non-memes.
The accuracy is greater than 0.97 on the test set.
Most predicted probabilities are either 0 or 1.
Based on this, I find that around 40\% of the images shared by the IRA accounts can be classified as memes (see Figure \ref{fig:classifier-percentage} for a histogram of predicted probability).

Figure \ref{fig:predicted-meme-nonmeme} in Appendix shows a sample of predicted meme vs. non-meme IRA images.
There is room for improvements in accuracy.
But since there is no universally accepted definition to serve as ground truth for labeling memes, this might be a simpler procedure without human labeling.
For better comparisons, the later analysis will only use the IRA subsample that I classified as memes (predicted probability >0.9) to compare with authentic memes on Reddit.
\subsection*{Extracting representations of visual features}
\label{sec:org680d515}

Most traditional image classification tasks are based on supervised learning, but it is hard to scale, especially for memes.
Another approach, recently picked-up by social scientists (eg, \citet{torres_2018_GiveMeFull}), is to extract
keypoints via scale-invariant feature transform (SIFT) and find Bag-of-Visual-Words (BOVW) feature representation \citep{sivic_2003_VideoGoogleText} by building patches around the neighbor of keypoints and finding clusters of patches.
However, this method tends to only focus on local features around the patches and can be not meaningful enough for interpretation.

\begin{figure}[!t]
    \includegraphics[width=.49\textwidth]{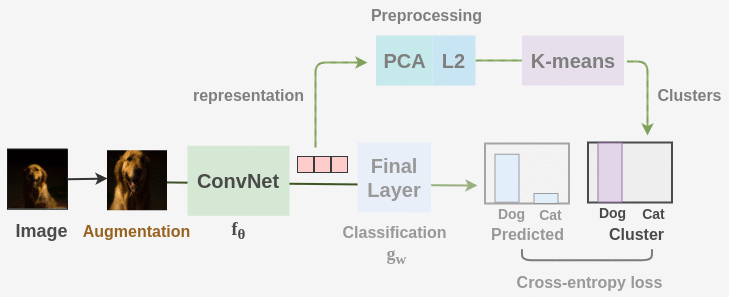}
    \caption{DeepCluster Pipeline (from \citet{chaudhary_2020_VisualExplorationDeepCluster})}
    \label{fig:deepcluster-pipeline}
\end{figure}

This paper leverages DeepCluster \citep[][also introduced to the social scientists by \citet{zhang_2021_ImageClusteringUnsupervised} that found successes in social science applications]{caron_2019_DeepClusteringUnsupervised}, a recent self-supervised method for clustering images.
See Figure \ref{fig:deepcluster-pipeline} for the pipeline for DeepCluster.
Specifically, DeepCluster learns \emph{pseudo-labels} iteratively by grouping features into clusters and uses the subsequent assignments as supervision to update the weights of the convolutional neural network (ConvNet).

\begin{figure}[!b]
    \includegraphics[width=.5\textwidth]{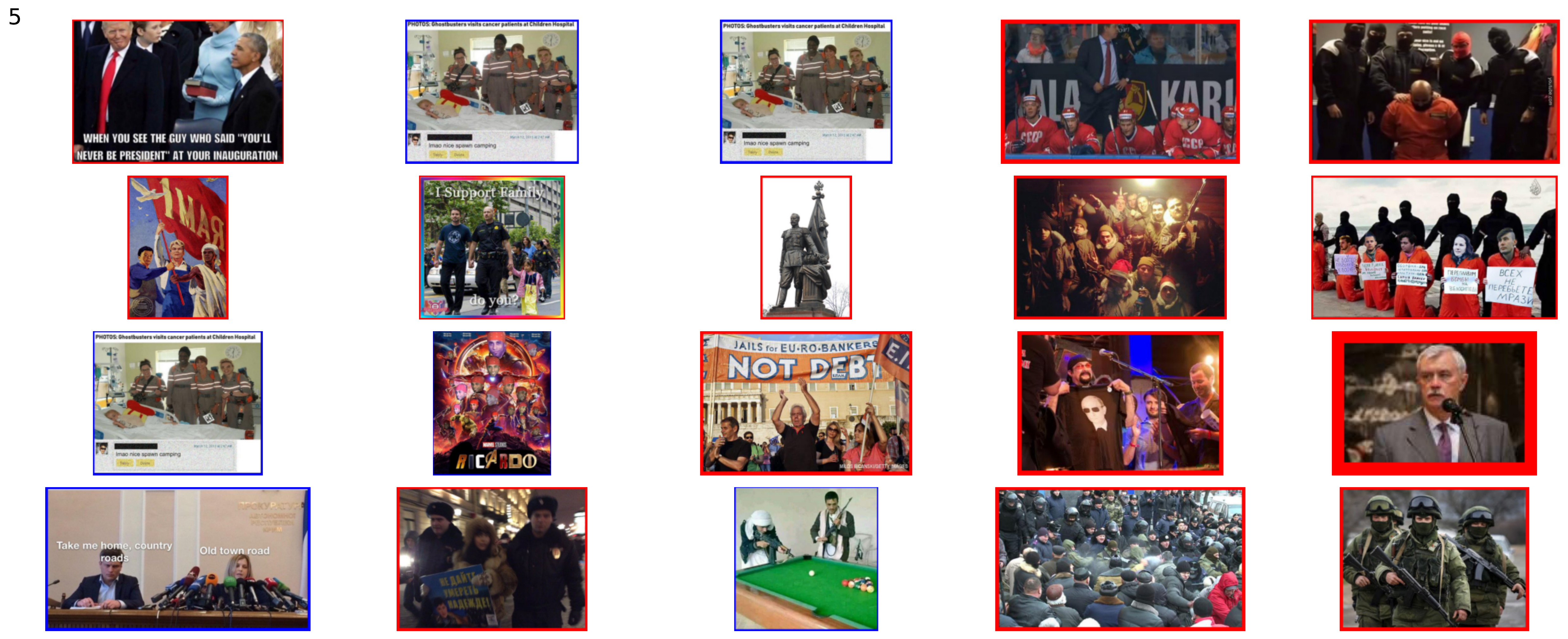}
    \rule[0.5ex]{\linewidth}{0.5pt}
    \includegraphics[width=.5\textwidth]{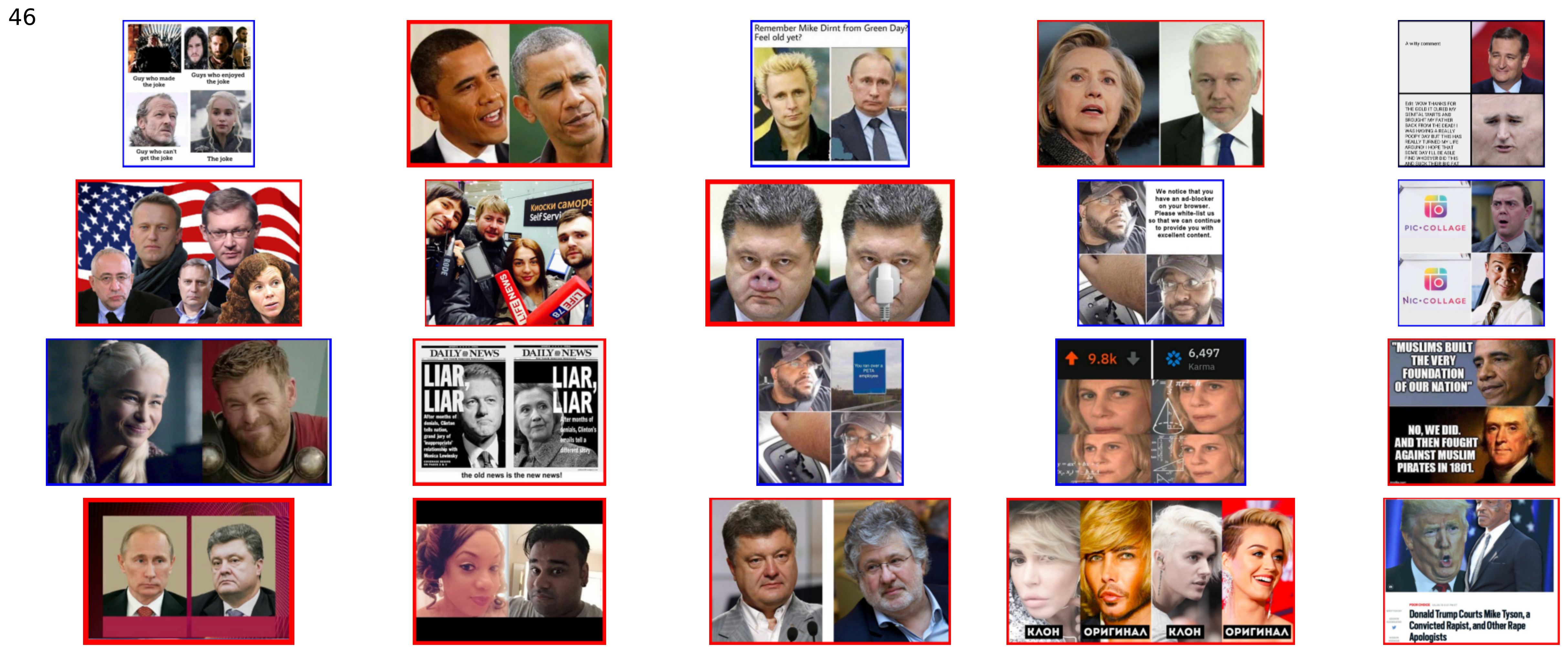}
    \caption{\textbf{Example memes from Clusters 5 (top) and 46 (bottom).}
    For each, the first row contains the 5 memes nearest to the center of the cluster;
    the 2-4 rows contain 15 random memes from that cluster.
    Red border indicates the meme is from IRA;
    blue border indicates the meme is from Reddit.
    }
    \label{fig:cluster_example}
\end{figure}

I feed a balanced sample of coordinated IRA memes and authentic Reddit memes (each of size 26K) into DeepCluster \emph{jointly}.
Notice that the IRA/Reddit labels are \emph{not} inputs of the model since this is an unsupervised algorithm, and we also don't want the model to memorize the labels at this stage.
For faster training, I use pre-trained weights released by the authors (based on the VGG-16 model trained on the ImageNet dataset).
ImageNet contains 1.28M images of 1000 categories such as scenes, places, and objects.
Thus, the learned embeddings should be helpful in finding general themes in images, not just localized features (e.g., textures) like those in BOVW.
The code and pre-trained weights are publicly available on the \href{https://github.com/facebookresearch/deepcluster}{GitHub repository of Facebook Research}.
I extract the final layer of the ConvNet before classification (a 4096-dimensional vector) for each meme as a representation of the visual features.

\subsection*{Cluster the memes and label the clusters}
\label{sec:org11d8e61}

\begin{figure}[!t]
    \vspace{-1cm}\hspace{-1.75cm}
    \includegraphics[width=.6\textwidth]{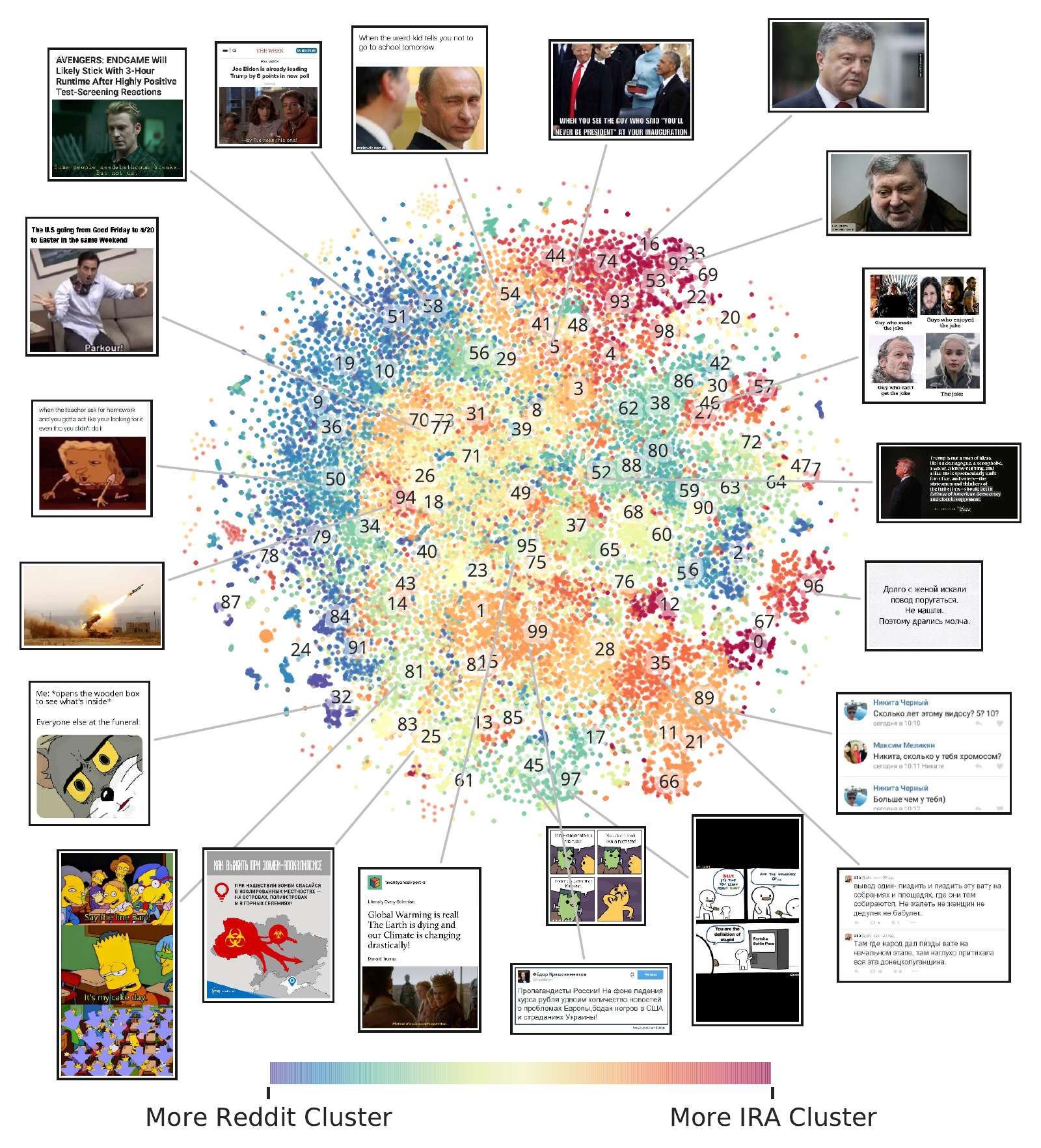}
    \caption{\textbf{t-SNE projection of IRA and Reddit memes.}
    Each point represents a meme from IRA or Reddit.
    Each color indicates a cluster (K-means with K=100 and Euclidean distance) on the embedding space learned from DeepCluster \citep{caron_2019_DeepClusteringUnsupervised}.
    Each number indicates the index of the cluster, labeled at the centroid of the cluster.
    Red indicates that the cluster has the highest percentage of memes from IRA;
    blue indicates that the cluster has the highest percentage of memes from Reddit.
    See Figure \ref{fig:cluster_share} for a complete list of the 100 clusters.
    }
    \label{fig:meme-cluster-tsne-rank-annotated}
\end{figure}

\begin{figure}[!t]
    \includegraphics[width=.44\textwidth]{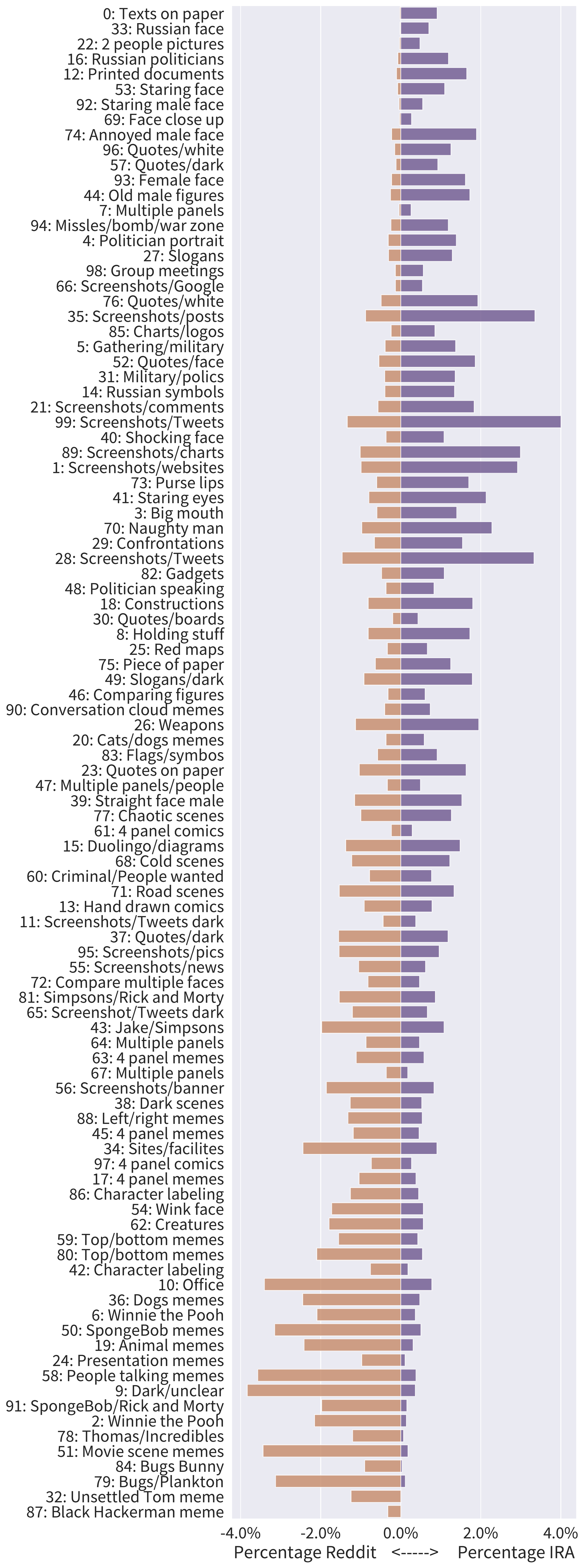}
    \caption{\textbf{Cluster labels and shares of IRA/Reddit memes}
    Percentages are calculated by the number of memes from IRA (Reddit) in a cluster out of total number of memes from IRA (Reddit), respectively. The clusters are ordered by the percentages.}
    \label{fig:cluster_share}
    \vspace{-1.5cm}
\end{figure}

After getting representations for each meme, I train the standard K-means algorithm with K=100 and Euclidean distance on the 4096-dimensional embedding space.
The choice of K is still arbitrary at this stage.
The idea is to choose a large enough K and combine similar clusters at the later stage.

After clustering, I see the images within each cluster to label the clusters.
Specifically, I sample 5 representative memes (memes that are nearest to the center of the clusters) and 15 random memes within that clusters (to ensure the robustness of the distance measure).
See Figure \ref{fig:cluster_example} for examples from Clusters 5 and 46.
Some more examples are in Figure \ref{fig:cluster_example_app} in the Appendix.
\section*{Preliminary Findings}
\label{sec:orgaac5fe6}

Figure \ref{fig:meme-cluster-tsne-rank-annotated} plots the t-SNE projection of the learned visual embeddings for each meme.
Each point represents a meme from IRA or Reddit;
each color indicates a clustering result from K-means.
Each number indicates the index of the cluster, labeled at the centroid of the cluster.
For each cluster, we also calculate the percentage of memes in that cluster (for IRA and Reddit memes separately).
Red indicates that the cluster has the highest percentage of memes from IRA;
blue indicates that the cluster has the highest percentage of memes from Reddit.

One can see that clusters located near the top of Figure \ref{fig:meme-cluster-tsne-rank-annotated} involve mostly pictures of public figures (politicians, celebrities).
Clusters located near the bottom involve mostly screenshots (Twitter/Facebook posts, quotes/slogans, news websites/headlines, messages, etc.).
Clusters located near the top left consist of the common ``memes'': pictures surrounded by text.
Clusters located near the bottom left consist of comics, maps, charts, etc.
Most clusters are complex mixing of images, text, pictures, screenshots, and comics.

Figure \ref{fig:cluster_share} presents a complete list of clusters, and labels, along with the percentage count within IRA/Reddit memes.
For example, cluster 99 (one of the clusters involving Screenshots/Tweets) accounts for 4\% of the IRA memes.
The top row indicates that the cluster has the highest relative percentage of IRA memes (around 1\% of IRA memes and no Reddit memes);
the bottom row indicates that the cluster has the lowest relative percentage of IRA memes (0.5\% of Reddit memes and no IRA memes).
One can see that, towards the top of the list (more common in IRA memes), there are more themes around pictures of public figures, quotes, slogans, screenshots, and scenes related to military or gender.
In comparison, towards the bottom of the list (more common in Reddit memes), there are more comics, cats/dogs, superheroes, and movie scenes.
Noticeably, Reddit memes usually evolve around fixated ``frames'' where free online meme-creating tools can help you create memes with the same frame without editing the whole meme yourself.
Although these tools are wildly available in the West, it seems like the IRA accounts are not utilizing these tools to generate memes with popular frames.

\begin{figure}[!t]
    \includegraphics[width=.35\textwidth]{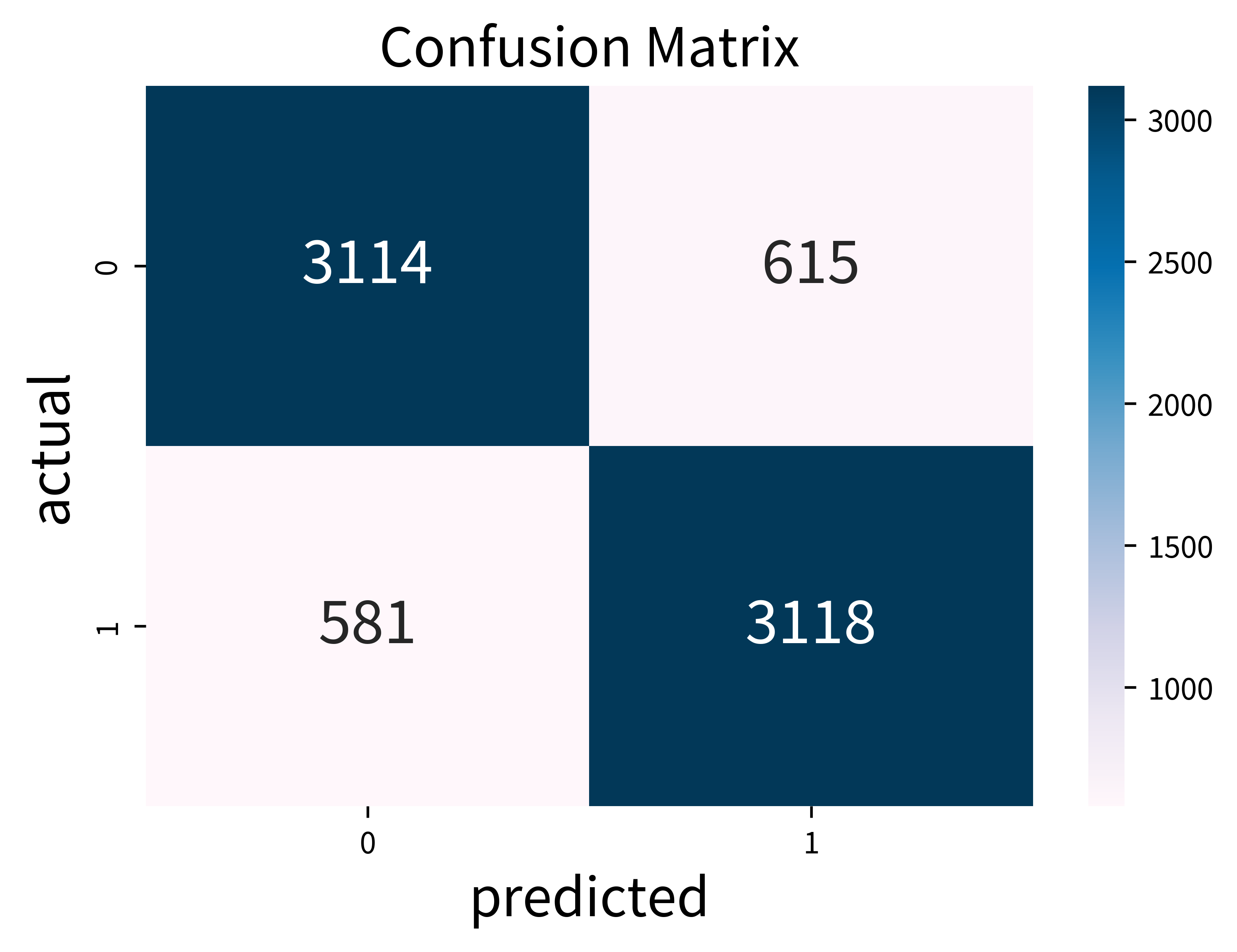}
    \caption{\textbf{Confusion matrix for logistic regression predicting IRA memes using only visual representations}}
    \label{fig:confusion_logistic}
\end{figure}


Can machine learning discern IRA memes from Reddit memes simply by using the 4096-dimensional visual representations?
I train a simple logistic regression using only the visual representations learned by DeepCluster with a 70/30 train/test split.
This simple baseline using visual representations alone achieves training accuracy 0.90, AUC 0.91, testing accuracy 0.84, precision 0.84, recall 0.84, F\textsubscript{1}-score 0.84.
See Figure \ref{fig:confusion_logistic} for the confusion matrix for this logistic regression.

\section*{Discussions and Future Steps}
\label{sec:org78dda5b}

In these preliminary experiments, I find that coordinated IRA memes and authentic Reddit memes share a large set of visual themes but with varying degrees.
IRA memes promote more pictures of celebrities, quotes, screenshots, and images related to military and gender.
Reddit memes involve more comics and movie characters.
I also find that using a simple logistic regression on the learned visual representations can reasonably discern coordinated memes from authentic ones.

The proposed method, based on DeepCluster \citep{caron_2019_DeepClusteringUnsupervised}, does not rely on labels and can learn broader themes of images.
In contrast, BOVW only learns about local visual features within patches, and pHash requires that images be nearly identical.
They can be less useful in identifying the visual themes of memes.

I plan to extend this framework to find better representations of visual themes:
\begin{itemize}
\item With the successes of multimodal transformer models (such as VisualBERT, ViLBERT, and VL-BERT) in \href{https://ai.facebook.com/blog/hateful-memes-challenge-winners/}{Facebook's Hateful Memes Challenge}, we can extract texts and entity/race tags and learn a more flexible model to get richer embeddings not only based on vision but also interacts with texts and other augmented information.
\item It is possible to better preprocess memes to strip off structures that are less related to themes (such as a number of panels within a meme) so that meme structures would not dominate during clustering.
\item It is also possible to utilize more flexible clustering models so that each meme does not only belong to one cluster but a distribution of clusters (similar to the Latent Dirichlet Allocation, \citet{blei_2003_LatentDirichletAllocation}) or even to include covariates such as source, time, or other metadata for building clusters (similar to the Structural Topic Model, \citet{roberts_2013_StructuralTopicModel}).
\end{itemize}

\appendix

\section*{Appendix}
\label{sec:orgc386169}

\begin{figure*}[t]
    \vspace{-1cm}
    \hspace*{-1cm}\includegraphics[width=.5\textwidth]{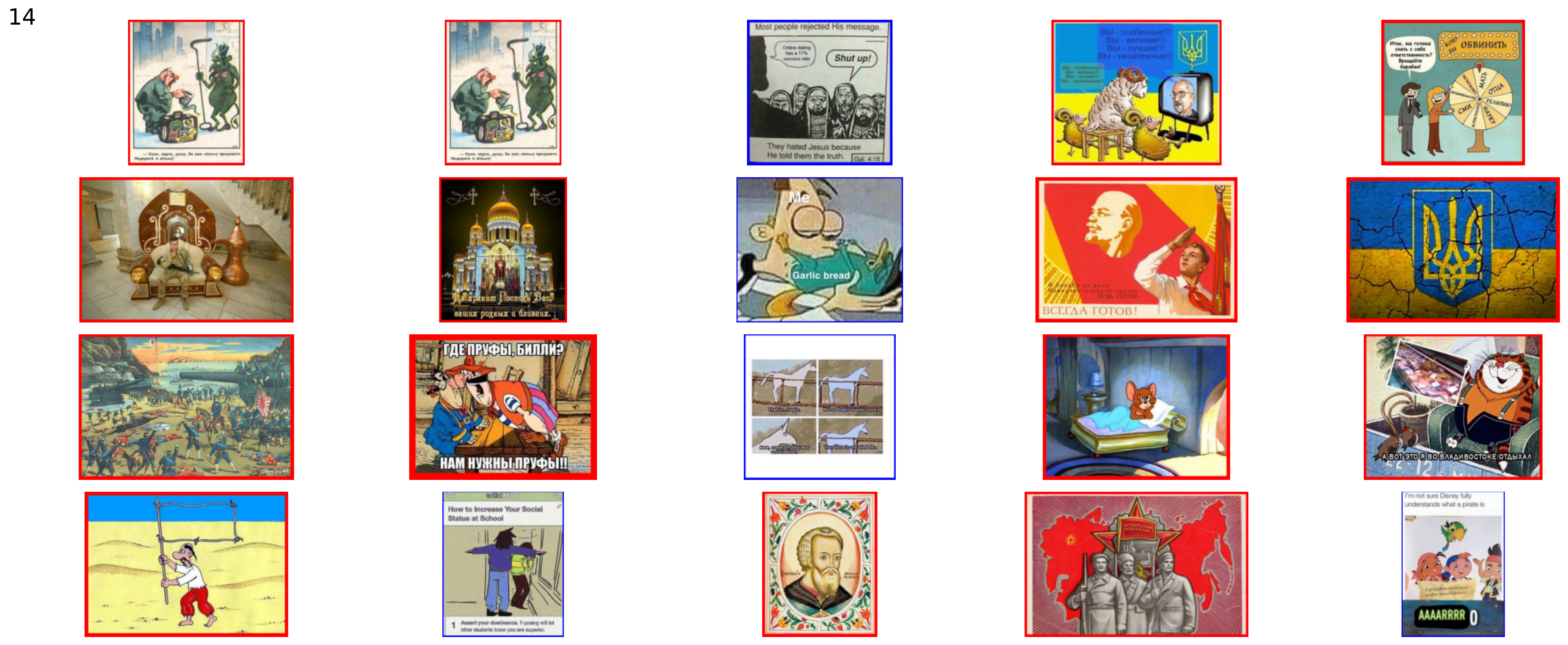}
    \hspace{1.5cm}
    \includegraphics[width=.5\textwidth]{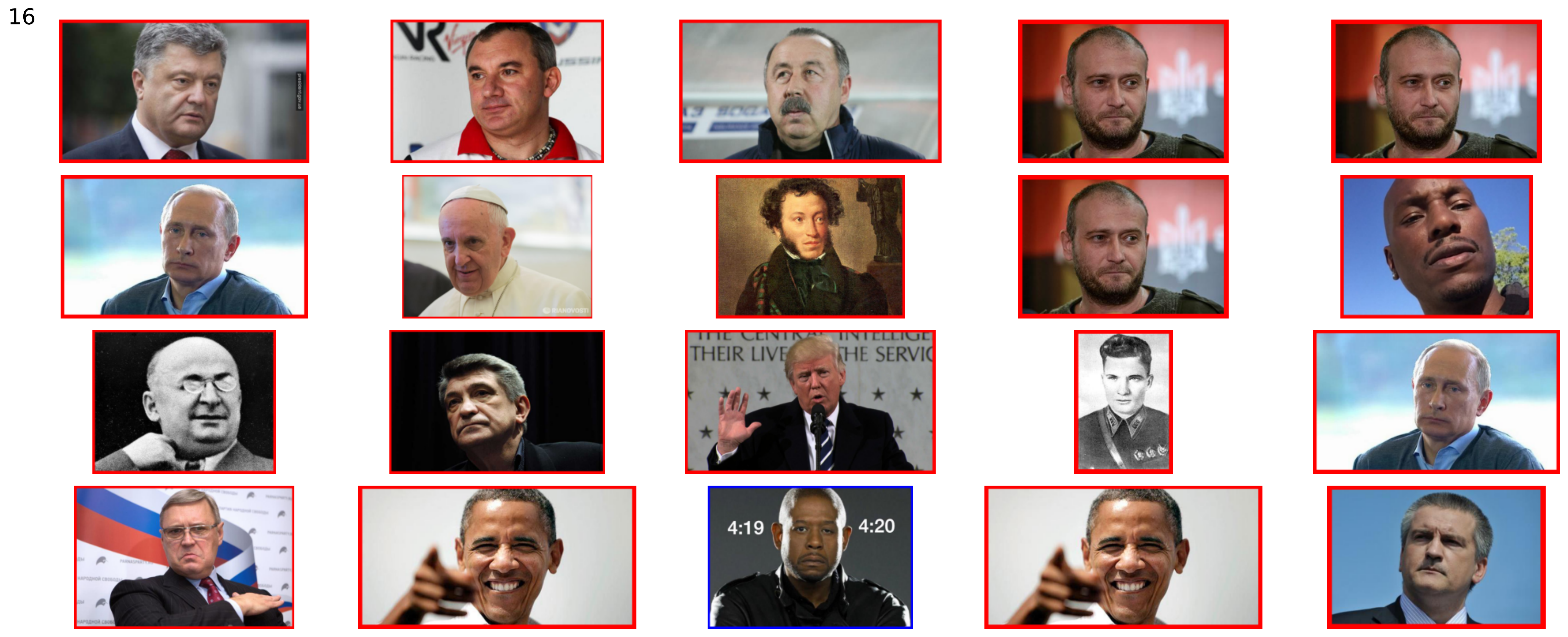}
    \rule[0.5ex]{\linewidth}{0.5pt}
    \hspace*{-1cm}\includegraphics[width=.5\textwidth]{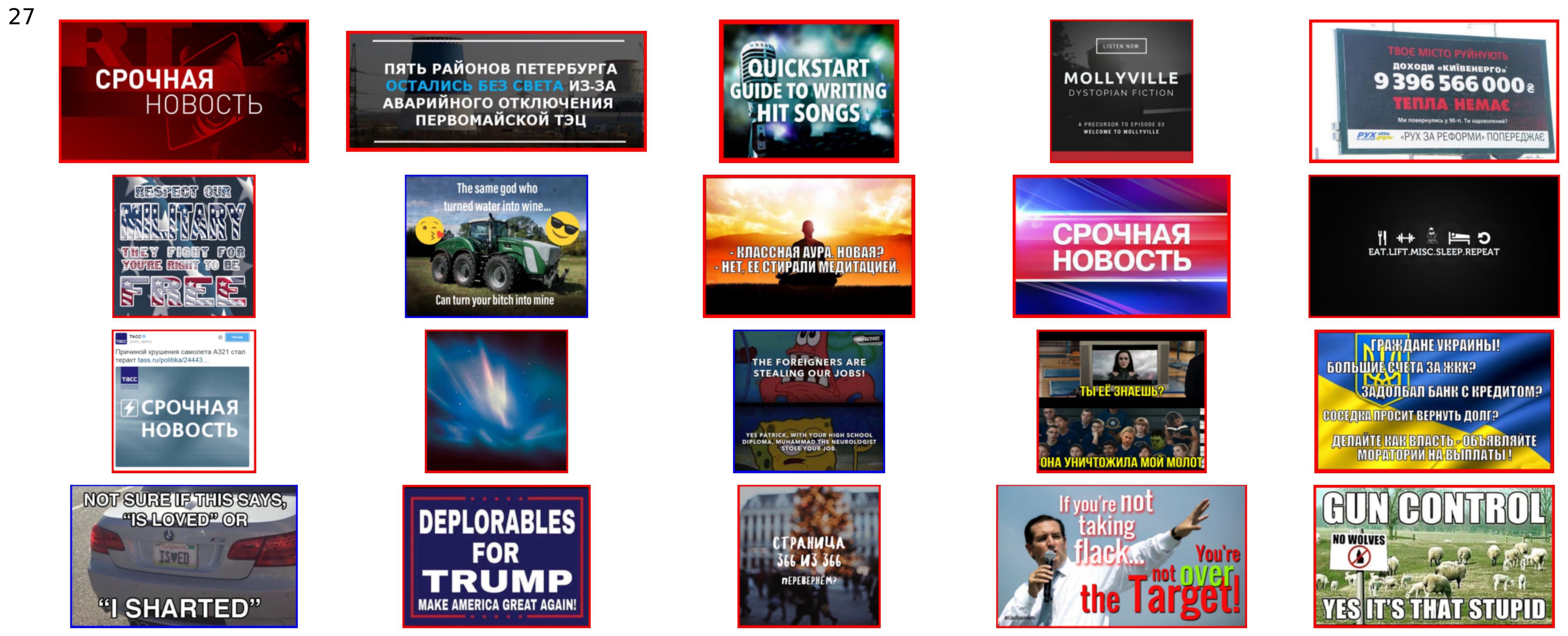}
    \hspace{1.5cm}
    \includegraphics[width=.5\textwidth]{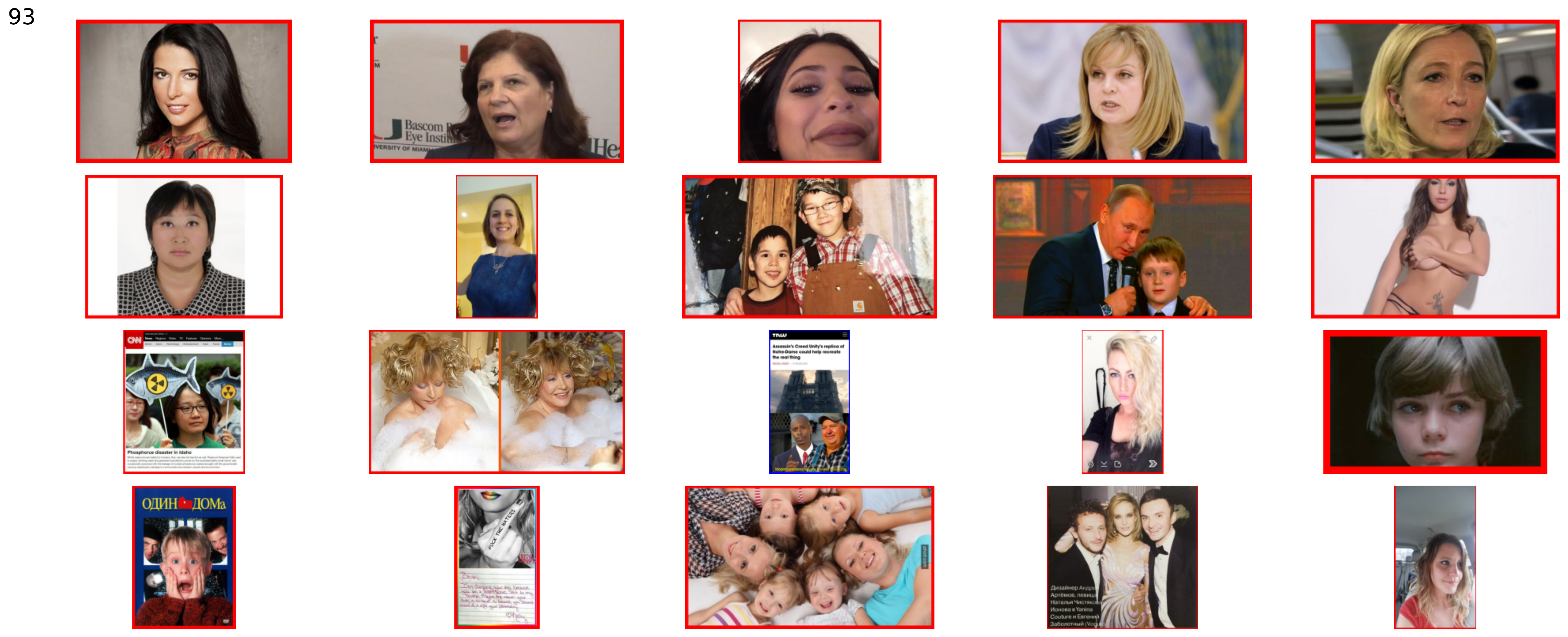}
    \rule[0.5ex]{\linewidth}{0.5pt}
    \hspace*{-1cm}\includegraphics[width=.5\textwidth]{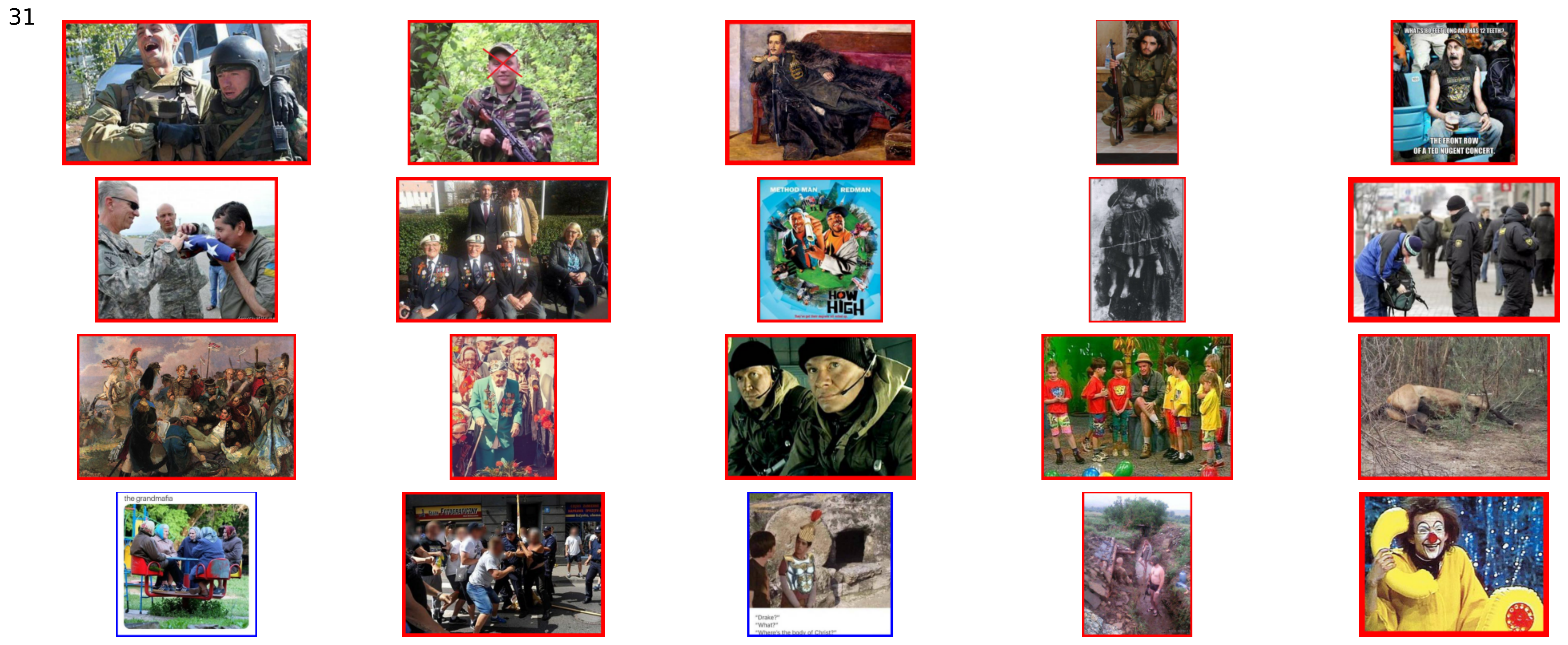}
    \hspace{1.5cm}
    \includegraphics[width=.5\textwidth]{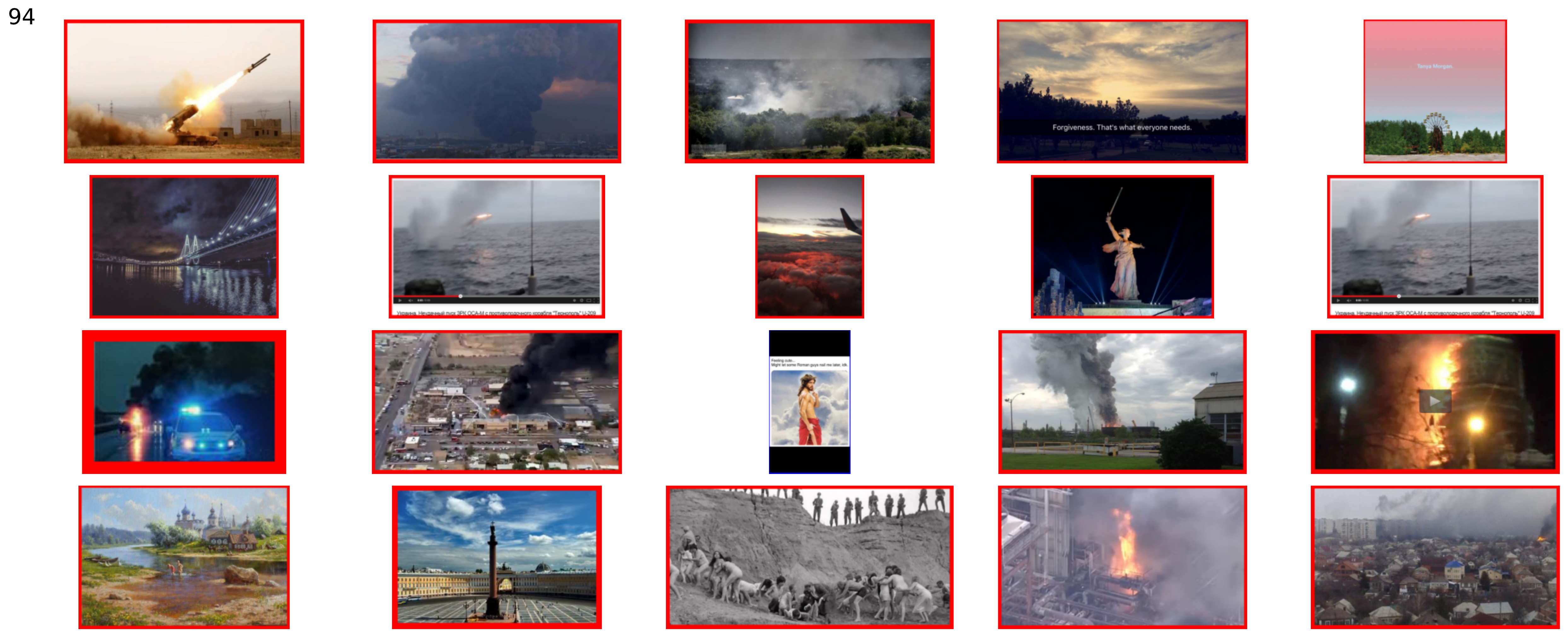}
    \rule[0.5ex]{\linewidth}{0.5pt}
    \hspace*{-1cm}\includegraphics[width=.5\textwidth]{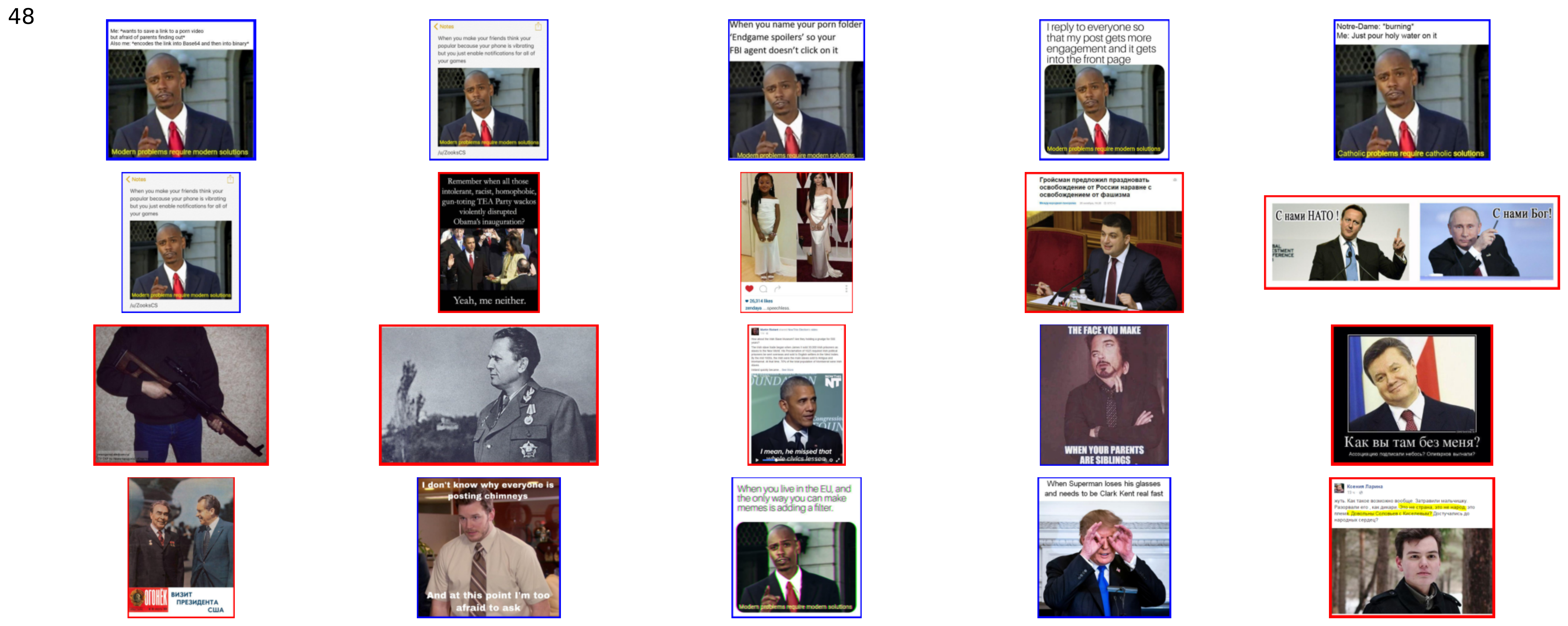}
    \hspace{1.5cm}
    \includegraphics[width=.5\textwidth]{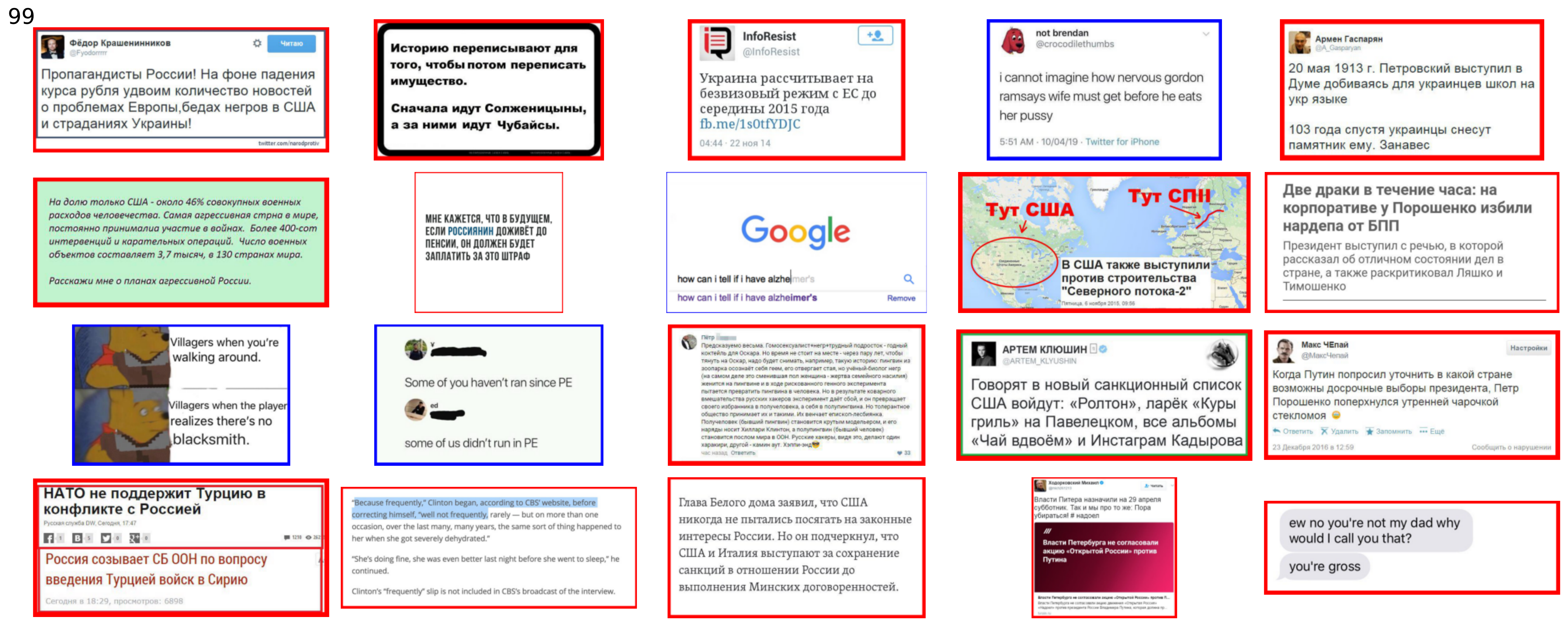}
    \rule[0.5ex]{\linewidth}{0.5pt}
    \hspace*{-1cm}\includegraphics[width=.5\textwidth]{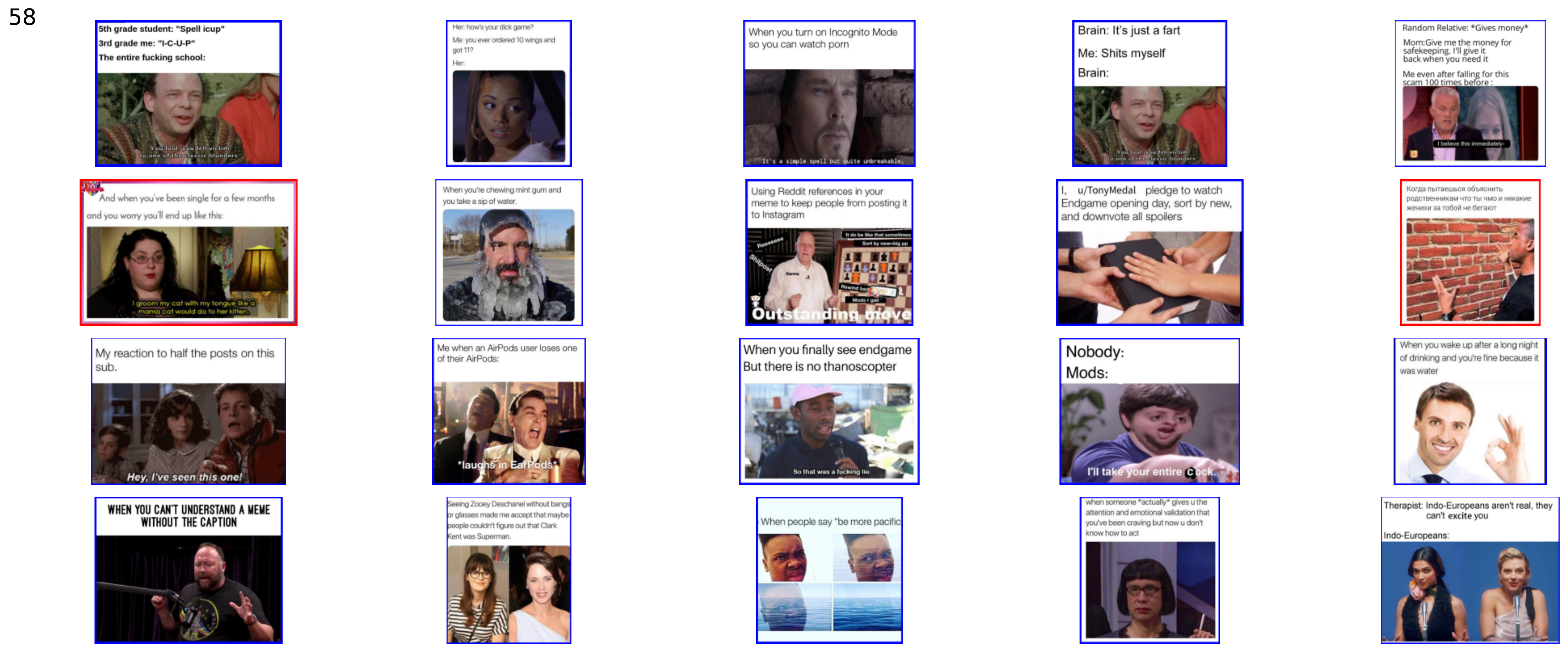}
    \hspace{1.5cm}
    \includegraphics[width=.5\textwidth]{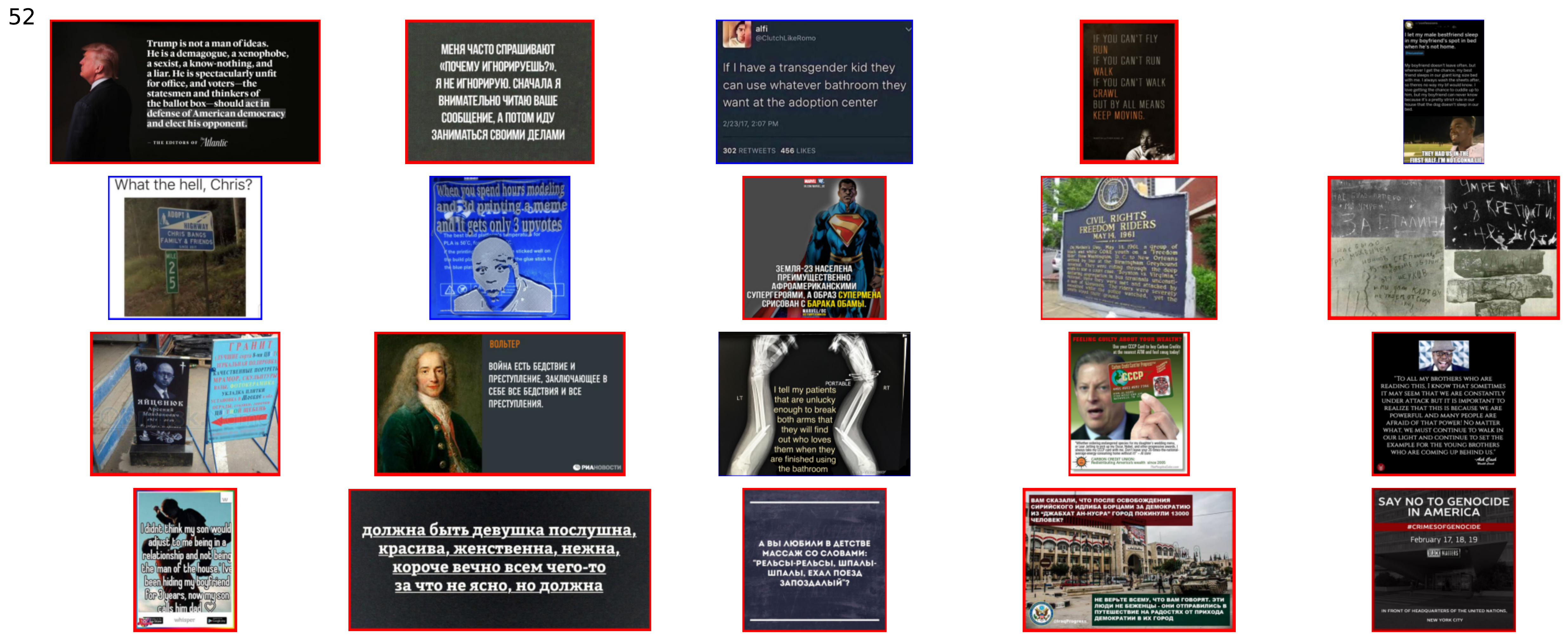}
    \rule[0.5ex]{\linewidth}{0.5pt}
    \hspace*{-1cm}\includegraphics[width=.5\textwidth]{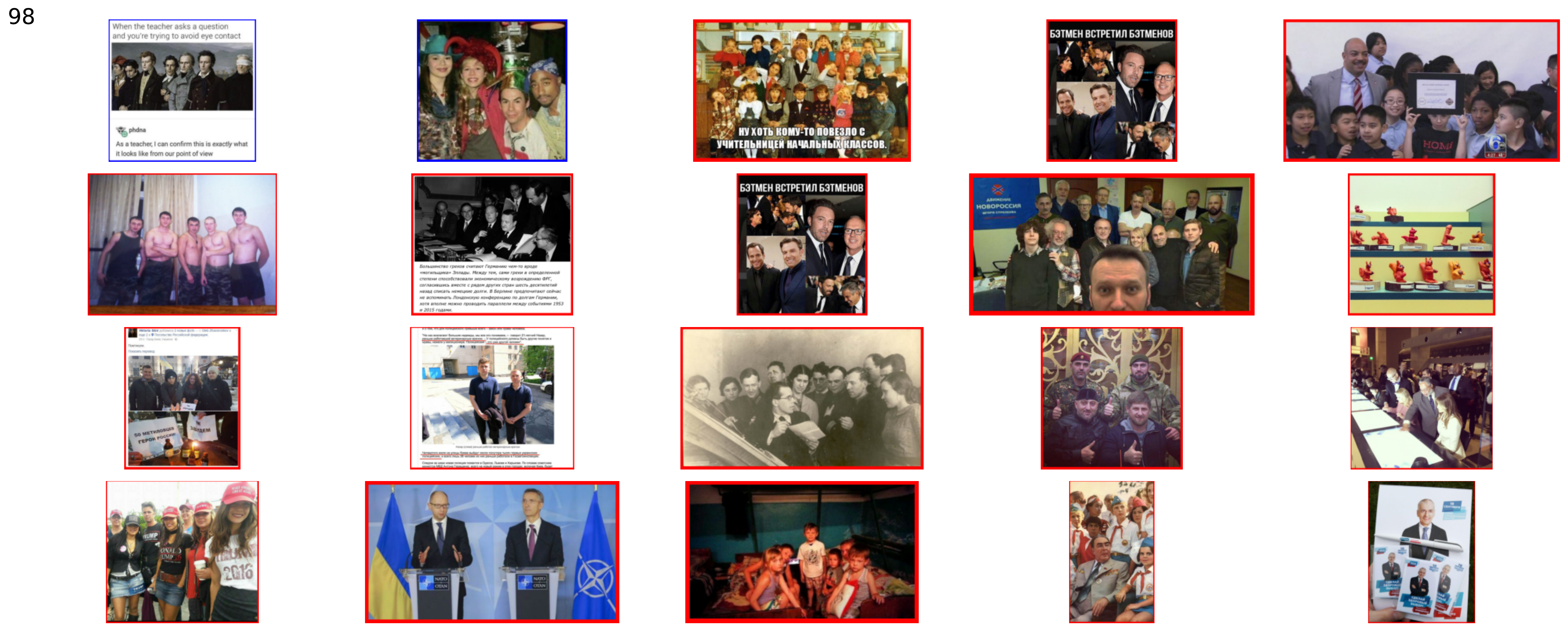}
    \hspace{1.5cm}
    \includegraphics[width=.5\textwidth]{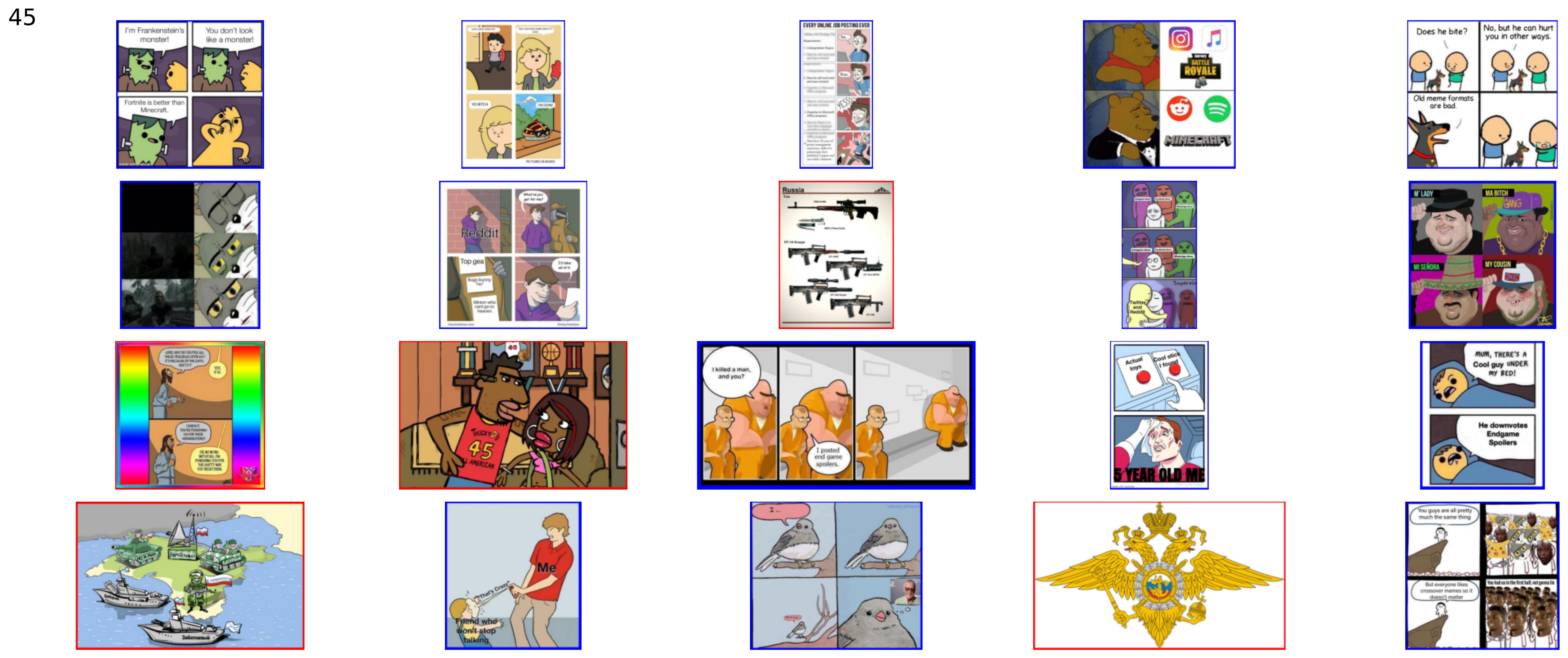}
    \caption{\textbf{Representative memes from selected clusters.}
    For each panel, the index of cluser is on the top left. The first row contains the 5 memes nearest to the center of the cluster;
    the 2-4 rows contain 15 random memes from that cluster.
    Red border indicates the meme is from IRA;
    blue border indicates the meme is from Reddit.}
    \label{fig:cluster_example_app}
\end{figure*}

\begin{figure*}[t]
    \centering
    \includegraphics[width=.65\textwidth]{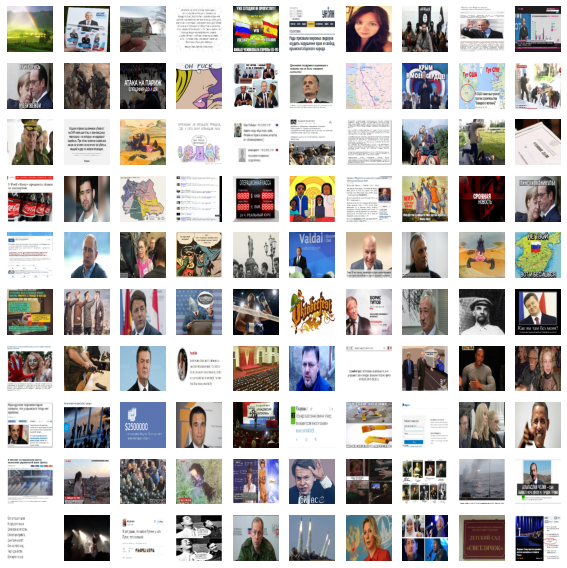}\\[2ex]
    \rule[2ex]{\linewidth}{1pt}
    \includegraphics[width=.65\textwidth]{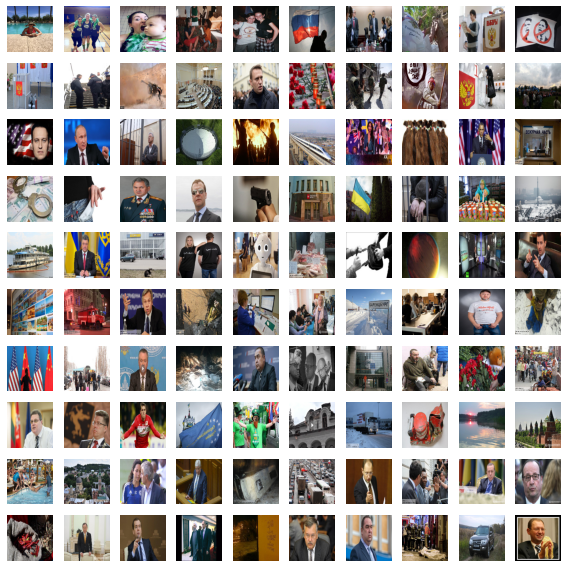}
    \caption{\textbf{Predicted memes (top) and non-memes (bottom) of IRA images.} See Methodology for details.}
    \label{fig:predicted-meme-nonmeme}
\end{figure*}

\begin{itemize}
\item Figure \ref{fig:cluster_example_app}: examples of representative memes from selected clusters.
\item Figure \ref{fig:predicted-meme-nonmeme}: examples of IRA images predicted as memes and non-memes.
\end{itemize}

\label{sec:org7b84d99}
\bibliography{../../Zotero/bib/all_papers_zotero}

\begin{thebibliography}{16}
\providecommand{\natexlab}[1]{#1}

\bibitem[{Alizadeh et~al.(2020)Alizadeh, Shapiro, Buntain, and
  Tucker}]{alizadeh_2020_ContentbasedFeaturesPredict}
Alizadeh, M.; Shapiro, J.~N.; Buntain, C.; and Tucker, J.~A. 2020.
\newblock Content-Based Features Predict Social Media Influence Operations.
\newblock \emph{Science Advances}, 6(30): eabb5824.

\bibitem[{Beskow, Kumar, and
  Carley(2020)}]{beskow_2020_EvolutionPoliticalMemes}
Beskow, D.~M.; Kumar, S.; and Carley, K.~M. 2020.
\newblock The Evolution of Political Memes: {{Detecting}} and Characterizing
  Internet Memes with Multi-Modal Deep Learning.
\newblock \emph{Information Processing \& Management}, 57(2): 102170.

\bibitem[{Blei, Ng, and Jordan(2003)}]{blei_2003_LatentDirichletAllocation}
Blei, D.~M.; Ng, A.~Y.; and Jordan, M.~I. 2003.
\newblock Latent Dirichlet Allocation.
\newblock \emph{The Journal of Machine Learning Research}, 3(null): 993--1022.

\bibitem[{Caron et~al.(2019)Caron, Bojanowski, Joulin, and
  Douze}]{caron_2019_DeepClusteringUnsupervised}
Caron, M.; Bojanowski, P.; Joulin, A.; and Douze, M. 2019.
\newblock Deep {{Clustering}} for {{Unsupervised Learning}} of {{Visual
  Features}}.
\newblock \emph{arXiv:1807.05520 [cs]}.

\bibitem[{Chaudhary(2020)}]{chaudhary_2020_VisualExplorationDeepCluster}
Chaudhary, A. 2020.
\newblock A {{Visual Exploration}} of {{DeepCluster}}.

\bibitem[{Cirone and Hobbs(2022)}]{cirone_2022_AsymmetricFloodingTool}
Cirone, A.; and Hobbs, W. 2022.
\newblock Asymmetric Flooding as a Tool for Foreign Influence on Social Media.
\newblock \emph{Political Science Research and Methods}, 1--12.

\bibitem[{DiResta, Grossman, and
  Siegel(2021)}]{diresta_2021_InHouseVsOutsourced}
DiResta, R.; Grossman, S.; and Siegel, A. 2021.
\newblock In-{{House Vs}}. {{Outsourced Trolls}}: {{How Digital Mercenaries
  Shape State Influence Strategies}}.
\newblock \emph{Political Communication}, 0(0): 1--32.

\bibitem[{Du, Masood, and Joseph(2020)}]{du_2020_UnderstandingVisualMemes}
Du, Y.; Masood, M.~A.; and Joseph, K. 2020.
\newblock Understanding {{Visual Memes}}: {{An Empirical Analysis}} of {{Text
  Superimposed}} on {{Memes Shared}} on {{Twitter}}.
\newblock \emph{Proceedings of the International AAAI Conference on Web and
  Social Media}, 14: 153--164.

\bibitem[{Garimella and
  Eckles(2020)}]{garimella_2020_ImagesMisinformationPolitical}
Garimella, K.; and Eckles, D. 2020.
\newblock Images and Misinformation in Political Groups: {{Evidence}} from
  {{WhatsApp}} in {{India}}.
\newblock \emph{Harvard Kennedy School Misinformation Review}.

\bibitem[{Roberts et~al.(2013)Roberts, Stewart, Tingley, and
  Airoldi}]{roberts_2013_StructuralTopicModel}
Roberts, M.~E.; Stewart, B.~M.; Tingley, D.; and Airoldi, E.~M. 2013.
\newblock The Structural Topic Model and Applied Social Science.
\newblock In \emph{Advances in Neural Information Processing Systems Workshop
  on Topic Models: {{Computation}}, Application, and Evaluation}.

\bibitem[{{Sivic} and {Zisserman}(2003)}]{sivic_2003_VideoGoogleText}
{Sivic}; and {Zisserman}. 2003.
\newblock Video {{Google}}: A Text Retrieval Approach to Object Matching in
  Videos.
\newblock In \emph{Proceedings {{Ninth IEEE International Conference}} on
  {{Computer Vision}}}, 1470--1477 vol.2.

\bibitem[{Torres(2018)}]{torres_2018_GiveMeFull}
Torres, M. 2018.
\newblock Give Me the Full Picture: {{Using}} Computer Vision to Understand
  Visual Frames and Political Communication.
\newblock \emph{Working Paper}, 30.

\bibitem[{Veit et~al.(2016)Veit, Matera, Neumann, Matas, and
  Belongie}]{veit_2016_COCOTextDatasetBenchmark}
Veit, A.; Matera, T.; Neumann, L.; Matas, J.; and Belongie, S. 2016.
\newblock {{COCO-Text}}: {{Dataset}} and {{Benchmark}} for {{Text Detection}}
  and {{Recognition}} in {{Natural Images}}.
\newblock \emph{arXiv:1601.07140 [cs]}.

\bibitem[{Zannettou et~al.(2018)Zannettou, Caulfield, Blackburn, De~Cristofaro,
  Sirivianos, Stringhini, and
  {Suarez-Tangil}}]{zannettou_2018_OriginsMemesMeans}
Zannettou, S.; Caulfield, T.; Blackburn, J.; De~Cristofaro, E.; Sirivianos, M.;
  Stringhini, G.; and {Suarez-Tangil}, G. 2018.
\newblock On the {{Origins}} of {{Memes}} by {{Means}} of {{Fringe Web
  Communities}}.
\newblock \emph{arXiv:1805.12512 [cs]}.

\bibitem[{Zannettou et~al.(2019)Zannettou, Caulfield, Bradlyn, De~Cristofaro,
  Stringhini, and Blackburn}]{zannettou_2019_CharacterizingUseImages}
Zannettou, S.; Caulfield, T.; Bradlyn, B.; De~Cristofaro, E.; Stringhini, G.;
  and Blackburn, J. 2019.
\newblock Characterizing the {{Use}} of {{Images}} in {{State-Sponsored
  Information Warfare Operations}} by {{Russian Trolls}} on {{Twitter}}.
\newblock \emph{arXiv:1901.05997 [cs]}.

\bibitem[{Zhang and Peng(2021)}]{zhang_2021_ImageClusteringUnsupervised}
Zhang, H.; and Peng, Y. 2021.
\newblock Image {{Clustering}}: {{An Unsupervised Approach}} to {{Categorize
  Visual Data}} in {{Social Science Research}}.

\end{thebibliography}
\bibliographystyle{aaai22}
\end{document}